\documentclass{sig-alternate}

\usepackage[ruled,vlined,noend]{algorithm2e}
\usepackage{graphicx}
\usepackage{url}
\usepackage[utf8]{inputenc}
\usepackage{balance}  % for  \balance command ON LAST PAGE  (only there!)
\usepackage[bf]{subfigure}
\usepackage{floatflt}
\usepackage{latexsym}
\usepackage{amsfonts}
\usepackage{amsmath}

\begin{document}

\title{Public Data Integration with WebSmatch}
\author{
\alignauthor
Remi Coletta$^{(a)}$, Emmanuel Castanier$^{(b)}$, Patrick Valduriez$^{(b)}$, \\
Christian Frisch$^{(c)}$, DuyHoa Ngo$^{(a)}$,  Zohra Bellahsene$^{(a)}$\\
(a,b) INRIA and LIRMM, Montpellier, France ~~~~~ (c) Data Publica, Paris, France \\
{\normalsize(a) \{FirstName.Lastname\}@lirmm.fr ~~~~~(b) \{FirstName.Lastname\}@inria.fr ~~~~~  (c) christian.frisch@data-publica.com}
}

\date{}
\maketitle

\begin{abstract}
Integrating open data sources can yield high value information but
raises major problems in terms of metadata extraction, data source
integration and visualization of integrated data.
In this paper, we describe WebSmatch, a flexible environment
for Web data integration, based
on a real, end-to-end data integration scenario over public data
from Data Publica\footnote{\url{http://www.data-publica.com}}. 
WebSmatch supports the full process of importing,
refining and integrating data sources and uses third party tools
for high quality visualization. We use a typical scenario of public data integration
which involves problems not solved by currents tools: poorly
structured input data sources (XLS files) and rich visualization of
integrated data. 
\end{abstract}

\section{Introduction}

%Recent open data government initiatives, such as \url{data.gov}
%or \url{data.gov.uk}, promote the idea that certain data produced by public organizations
%should be freely available to everyone to use and republish as they wish.
%As a result, a lot of open data sources are now available on public
%organization's web sites, in various formats.
%Therefore, integrating open data sources from different organizations
%can yield high value information. For instance, matching gas emission
%data with climatic data for a given country or city can be valuable to
%better understand pollution.
%%
%Based on this observation, Data Publica, a french company, 
%provides added value over the public data sources they crawl,
%such as visualization of data sources or production of integrated
%data.
%Achieving this goal raises the followings problems:

Recent open data government initiatives, such as  \url{data.gov},  \url{data.gov.uk},  \url{data.gouv.fr} promote the idea that certain data produced by public organizations should be freely available to everyone to use and republish as they wish. As a result, a lot of open data sources are now available on public organization’s web sites, in various formats. 

Integrating open data sources from different organizations can yield high value information. For instance, matching gas emission data with climatic data for a given country or city can be valuable to better understand pollution. This rich local and targeted pool of information can also be leveraged to build new innovative services or, as a new source of business intelligence, to put in perspective business information with data such as weather, traffic, density of economic activities or touristic information in order to better understand current market dynamics and adapt product and services.

A large share of the available open data comes from large institutions
(such as Eurostat, World bank, UN....) using structured data formats
such as SDMX for statistical datasets or RDF for linked open
data. However, the majority of the data that can be found on open data
portals is available as unstructured data (such as spreadsheets). To
integrate these sources or deliver data that web applications or
mobile services can leverage, raw open data files must be structured
through a processing workflow and delivered through APIs (Application
Programming Interfaces). This workflow will ultimately transform ``human usable information'' such as spreadsheets into ``computer usable data'', drastically increasing the value of the open data published.

Based on this observation, Data Publica, a french company, provides added value over the public data sources they crawl, such as visualization of data sources or production of integrated data. Achieving this goal raises the followings problems:

\textbf{Metadata extraction.} Although structured formats exist to share and publish data, most of
the public data available on the Web are Excel spreadsheets, with no
difference between data and metadata. Detecting the metadata in such
data sources is a mandatory step before performing data integration.
To address this problem,  we exploit computer vision techniques to deal with complex tabular representations of spreadsheets
and machine learning techniques that take advantage of past human effort to automatically detect metadata in the next spreadsheets.

\textbf{Data sources integration.} 
In order to produce added value information over the public data sources, it is
necessary to integrate data sources together. 
For this purpose, we need to perform schema matching, in order to match
metadata structures \cite{bellahsene:2011}. In the context of open data, schema matching is
harder than in traditional data
integration in distributed database systems \cite{ozsu:2011}, mainly because important metadata
which are considered as implicit  by document's authors, are simply missing.
%data comes
% Zohra, Rémi: Ajouter une phrase pour dire en quoi le pb est nouveau
% par rapport à l'intégration de données classique en introduisant
% aussi et surtout le terme ``schema matching''
%For this purpose  needed to be matched, this is 
%schema matching. \textbf{ICI}
%
In terms of matching capabilities, we rely on  YAM++
\cite{YAM++}, a powerful tool for schema matching and ontology
alignment\footnote{YAM++ was recently ranked first
  at the Conference track of the OAEI
  competition over 15 participants. See the results at
  \url{http://oaei.ontologymatching.org/2011/} for more
  details.}.

\textbf{Visualization.} 
To ease users's access to public data requires visualizing 
with high quality graphical representation. 
In Data Publica, the visualization task is delegated to Google Data Explorer, 
a powerful collection of visualization tools.
However, Google Data Explorer imposes strict restrictions on input formats, such as separating 
data and metadata into different files and labeling metadata with some Google predefined concepts.
Therefore, using Google Data Explorer requires metadata extraction and integration as preliminary steps.

To perform these tasks, Data Publica uses WebSmatch  \url{http://websmatch.gforge.inria.fr/} , an environment
for Web data integration with a service-oriented
architecture with much flexibility for users and developers.
Most tools for metadata integration are implemented as heavy
clients and hard-coded with their own graphical interfaces.
They need to be downloaded and installed, which make them
hard to use with other independent tools (even if sources
are provided) and reduce their dissemination. In contrast,
WebSmatch is an open environment to be used as a Rich Internet
Application (RIA). 

In this paper, we describe the architecture  of WebSmatch based
on a real-life, end-to-end data integration scenario over public data
from Data Publica.

The paper is organized as follows. 
Section \ref{sec:scenario} introduces the motivating example in terms
of inputs (poorly structured files)
and outputs (rich visualization of integrated data).
Section \ref{sec:process} describes the data integration process with WebSmatch. 
Section \ref{sec:demo} presents WebSmatch metadata detection and integration services 
through the motiving example.
Section \ref{sec:related} discusses related work.
Section \ref{sec:conclu} concludes.

\section{Motivating example}
\label{sec:scenario}

In this section, we describe a real example  %demonstration scenario 
by giving the
inputs and outputs of the data integration process with WebSmatch.

Data Publica provides more than 12 000 files of public data. \cite{sizepublicdata}
However, even though data formats become richer and richer
in terms of semantics and expressivity (e.g. RDF), most
data producers do not use them much in practice, because
they require too much upfront work, and keep using simpler
tools like Excel.
As an example,  Data Publica has started
to crawl public data available from the French administration,
and found only 369 RDF files, compared with
148.509 .xls files. 
Unfortunately, no integration tool is able
to deal in an effective way with spreadsheets. As far as we
know, only two recent initiatives, OpenII \cite{openii} and Google Refine
\footnote{\url{http://code.google.com/p/google-refine/}} deal with Excel files. 
However, their importers are
very simple and make some strict restrictions over the input
spreadsheets. For instance, they require to have exactly one
table per sheet and all the attributes have to be in columns,
at the first line of the sheet. Unfortunately, people do not use
Excel in such proper way. And these importers proved to
be useless on real spreadsheets from Data Publica.
Thus, extracting metadata from such sources remains an open problem \cite{Hermans2010}.
To illustrate this problem in the remaining part of the paper, we use the following
spreadsheet files as input.

\begin{figure}[h]
     \includegraphics[width=8cm]{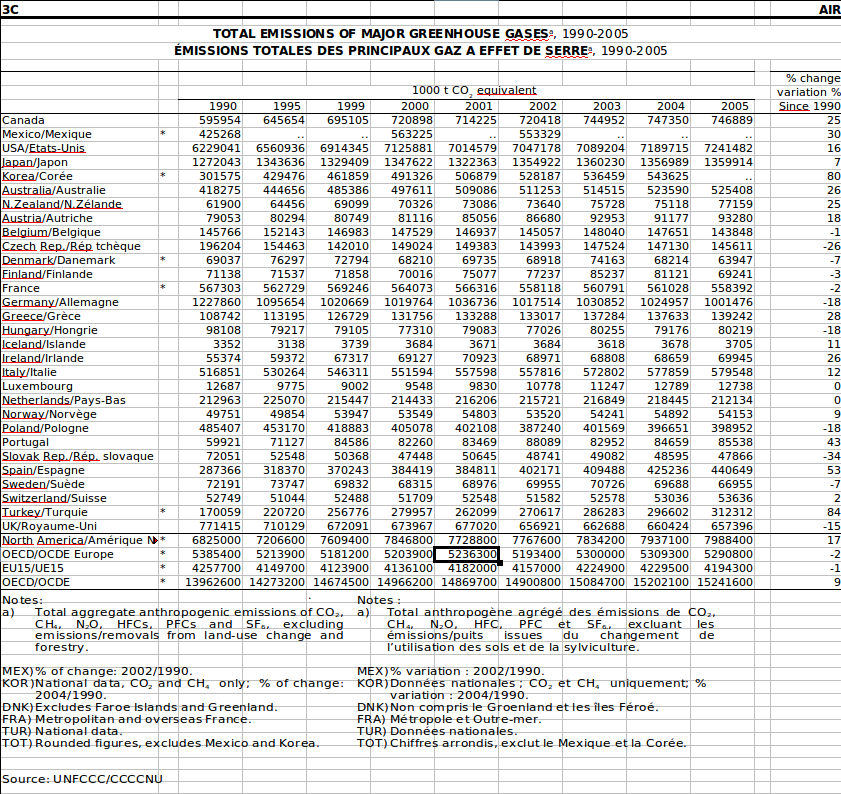}
     %\vspace{-0.5cm}
     \caption{Excel file crawled from OECD}
     \label{fig:xls}
\end{figure}

\subsection*{Input files}
For simplicity purposes, the scenario of this example involves only 2 data sources. To be representative of real-life public data, we choose two spreadsheet files:

%\begin{description}
%\item[Évolution de la température moyenne mondiale sur la période 1850-2008]
%\item[\url{http://www.data-publica.com/publication/1341}]
%\noindent
\url{http://www.data-publica.com/publication/1341} is an Excel file.
It contains data from  the Climatic Research Unit (\url{http://www.cru.uea.ac.uk/} )
about the temperature evolution in the world over the last decades.
This file is quite well formed, it only contains some blank lines and comments.

%\item[Statistiques sur l'air et le climat dans les pays de l'OCDE de 1980 à 2005]
%\item[\url{http://www.data-publica.com/publication/4736}]
%\noindent
 \url{http://www.data-publica.com/publication/4736}
is the Excel file depicted in Figure \ref{fig:xls}.
It contains data from OECD ( \url{http://www.oecd.org/} ) about gas emissions in the world. 
The file contains the evolution on the last 20 years on several countries and 4 OECD geographic zones\footnote{See \url{http://stats.oecd.org/glossary/} for more details about these zones.}.
This spreadsheet is much more complex: it involves several sheets, with several tables
per sheet. It contains several blank lines and comments, making it hard to automatically
detect the table.
In addition, it involves bi-dimensional tabular structures
%PV: bi-dimensional tabular structures? (il manque un mot je pense)
(Figure \ref{fig:xls}) and some column names
are missing. For instance, the author of this file probably has in mind that the line containing $\{1995, 2000\}$ 
should be labelled by "year", which is not obvious in the context of automatic integration.
%\end{description}

\subsection*{Expected results}

\begin{figure}[h]
\centering
     \includegraphics[width=8.6cm]{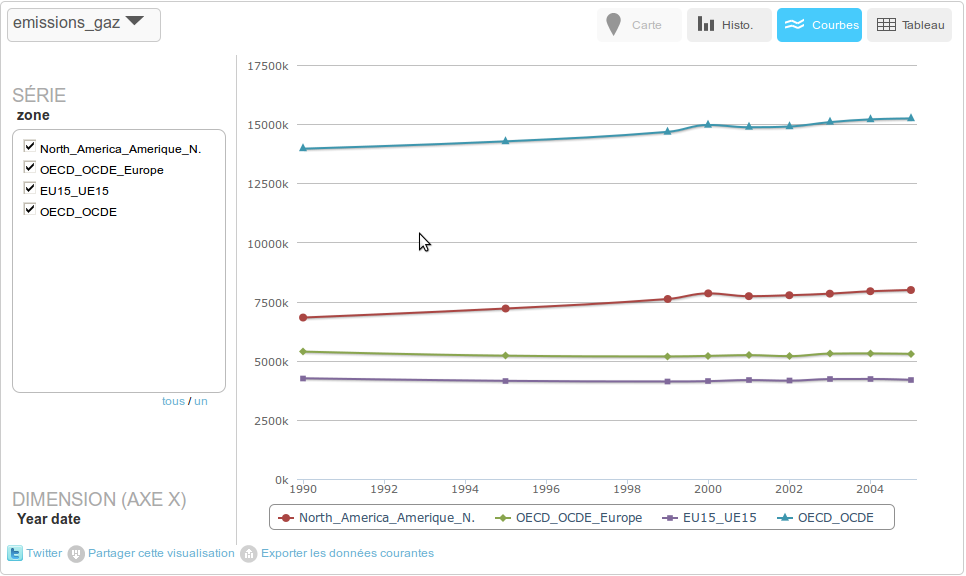}
     \caption{Evolution of gas emission}
     \label{fig:charts}
\end{figure}

\begin{figure}[h]
     \includegraphics[width=8cm]{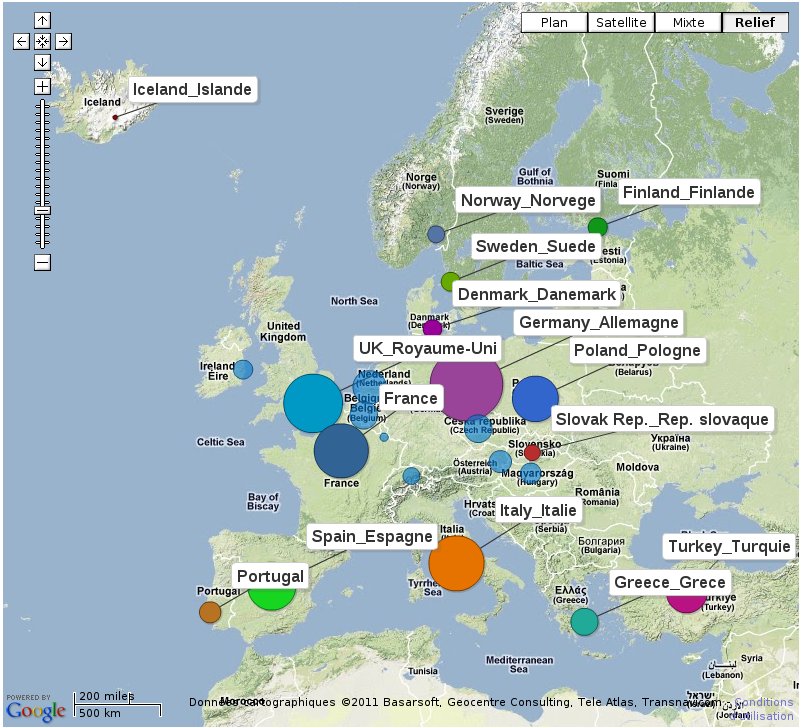}
     \caption{Geographic visualisation}
     \label{fig:map}
\end{figure}

\begin{figure}[h]
     \includegraphics[width=8cm]{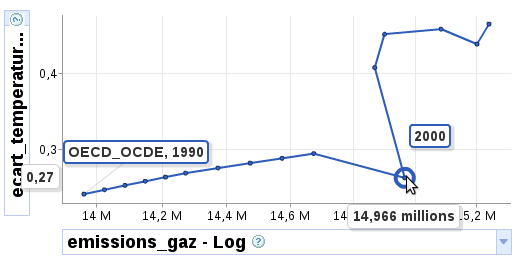}
     %\vspace{-0.5cm}
     \caption{Chart of integrated data}
     \label{fig:evolution}
\end{figure}

Charts (Figures \ref{fig:charts} and \ref{fig:evolution}), maps (Figure \ref{fig:map}) and additional
animations with timelines are visualizations obtained after extraction
of metadata and integration of 
the inputs described above.

%Déporté ici pour qu'il s'affiche page 3
\begin{figure*}[t]
\centering
     \includegraphics[width=15cm]{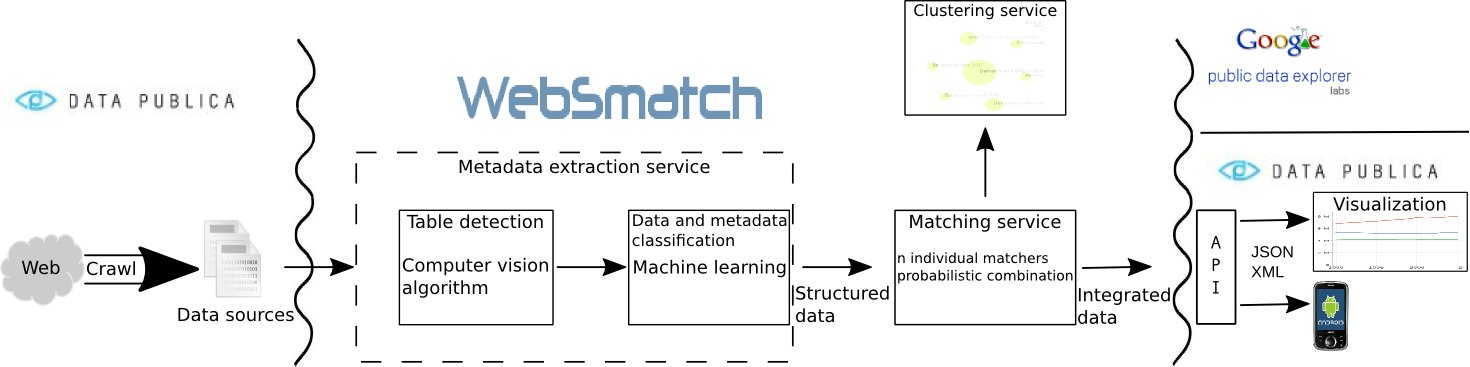}
     \vspace{-0.2cm}
     \caption{Data Integration process}
      \vspace{-0.2cm}
     \label{fig:flow}
\end{figure*}

Figure \ref{fig:charts} shows clearly that the emission of gas grows up significantly since 2000
in North America.
Since then, EU15 countries stabilized their emissions, which
corresponds to the Kyoto agreement.
Figure \ref{fig:map} is a screenshot of an animation of the same data on a map.

Figure \ref{fig:evolution} is a diagram involving both data sources. 
It correlates the evolution of temperature in the world with gas emission. 
Thus, it requires to integrate both data sources together.
The result shows clearly that the acceleration of the augmentation of temperature at the world level
increases significantly since 2000 with gas emission. 

To perform visualization, Websmatch exports the integrated data in
Data Set Publishing Language (DSPL) format \url{https://developers.google.com/public-data/}) . DSPL is used by Google Public Data Explorer and Data Publica's own API and visualisation engine.
%In terms of visualization, WebSmatch relies on Google Data Explorer, which
%ts own language as input, namely DSPL (Data Publishing Language).
Such format assumes the input data source to be precisely described. 
In particular,  data and metadata need be distinguished. 
The metadata (source, title, author, header) are described in an XML
file whereas the data are in Comma-Separated Values (CSV) files. In addition, metadata need to be tagged by some DSPL predefined concepts (hierarchy including times or geographical entities).
Such format is too strict to be usable by a large public, and quite difficult to manipulate, even for
computer scientists. 
Thus, although Google Data Explorer provides a powerful
collection of visualization tools, it  
requires much upfront work from the user, in particular, with public
spreadsheets like the ones
described above.

\section{Data Integration process}
\label{sec:process}

WebSmatch is a Rich Internet Application (RIA), meaning that Data
Publica is able to use it remotely, as a Web service, without any installation. 
To use all the WebSmatch components (integration, matching, clustering and export), 
Data Publica  simply needs to put some redirection from their back office.
The integration of WebSmatch and Data Publica is depicted in Figure \ref{fig:flow}.
It involves the following flow:\\

%\begin{description}
%\item[Crawling.] 
\noindent \textbf{Crawling.} Data Publica has developed a crawler dedicated to public data sources. It extracts data sources in various formats (such as Excel spreadsheets, ontologies, and XML files).
%DataPublica Feb 27
Data sources that are already well structured are directly transformed into DSPL and loaded into Data Publica's database. 
The other sources are sent to Websmatch (about 64\% of the volume)\\% and sends them WebSmatch.\\

%\item[Metadata extraction.] 
\noindent \textbf{Metadata extraction.}
The metadata extraction service takes as input raw files and extracts metadata to distinguish 
data from metadata. 
In the case of spreadsheets (more than 95 \% of public data), since
spreadsheet users often put several tables per sheet in their
document, the first task is to identify the different tables. This is achieved by a computer vision algorithm.
Then the different tables that have been identified are sent to the metadata classification service, which relies on Machine Learning techniques.\\

%\item[Matching]
\noindent \textbf{Matching.}

%Chopper du texte

As soon as the data sources have been cleaned, and data and metadata distinguished,
the data sources are ready to be matched.
This matching task achieves two goals. First, matching data sources
together allows discovering the overlaping between sources,
which can be exploited to generate integrated data.
%Second, matching with DSPL concepts makes the use of Google Data Explorer a lot easier.\\
%Modif 27feb
Second, concepts are identified in order to generate the appropriate
data description based on shared DSPL concepts defined with Data
Publica.
 %(ca rend le truc extensible et générique, pas besoin de mention de GPDE à ce moment).

%\item[Clustering] 
\noindent \textbf{Clustering.}
To deal with high numbers of data sources, as in Data Publica,
the usual 2-way matching approach (which makes visualization easy) becomes irrelevant.
Instead, we propose a schema clustering approach to visualize semantic similarities between sources.
Furthermore, clustering is a very intuitive way to perform recommendation to a user, who is looking for
related data sources.\\

%\item[Visualization] 
\noindent \textbf{Visualization.}
%Modif 27feb
Once data is extracted from the source file,  metadata is identified and concepts are matched, the information is structured as DSPL and exported.
The DSPL file is then loaded in Data Publica's database and served
through a generic API. This API supports different output formats such
as XML, CSV or Java Script Object Notation (JSON) and has filtering capabilities with standard functions (such as equals, greater than, in...) or geographic filters. This API is currently used by mobile applications and by Data Publica's own data visualization tool to display graphs, maps and compare datasets.
Alternatively, the DSPL file can be visualized in Google Data Public Explorer.\\
%Visualization is the end of the process. Data have been extracted from the web, data and metadata have been identified, and metadata have been matched. Data sources and integrated data are now ready to be visualized.
%\end{description}

\section{Running the complete workflow}
\label{sec:demo}

We now illustrate the complete workflow of using WebSmatch by Data Publica on
the scenario described in Section \ref{sec:scenario}. In order to couple the Data Publica back
office and the WebSmatch application, Data Publica uses WebSmatch services via its Crawl application. %(Figure \ref{fig:crawl}). 
Using the option "Publish (WS)" on its application redirects the crawled document to WebSmatch and the Data Publica user is also redirected to the WebSmatch editor main frame.

\begin{figure}[h]
     \includegraphics[width=8.7cm]{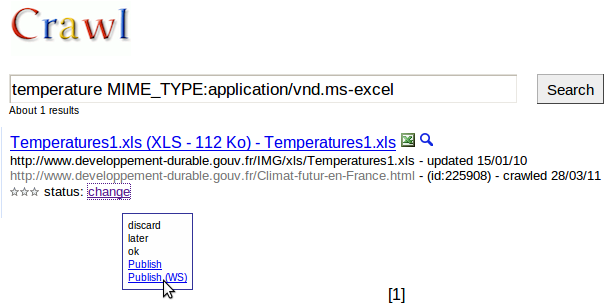}
     \caption{Data Publica Crawl application}
     \label{fig:crawl}
\end{figure}

\subsection*{Metadata Detection} 

After the Crawl (see Figure \ref{fig:crawl}), 
the user is redirected to
the WebSmatch RIA. It is important to note that Excel files (such as
.xls, for which there is no XML version) are not structured at all. As can be
seen in Figure \ref{fig:xls}, they can contain lots of artifacts such as
blank lines, blank columns, titles, comments, and not only a simple table. 

To get all the metadata and data, the chosen file is parsed and then,
two processes are applied to it. The first process relies on a
combination of computer vision algorithms.

Using the jexcelapi\footnote{\url{http://jexcelapi.sourceforge.net/}} library as a wrapper,
the spreadsheet is first translated into a 0/1 bitmap (0 for void cell / 1 for non empty cell).%, in which color are datatype of the cell.
%PV: je ne comprend pas le bout de phrase précédent

%To achieve this task, the spreadsheet is first translated in a Black and White bitmap, 
%using the jexcelapi\footnote{\url{http://jexcelapi.sourceforge.net/}} library as a wrapper.
In this bitmap, we run a connected component detection algorithm.
Algorithm \ref{algo:cc} takes as input a function indicating the color of a point in a bitmap
(in our case, a datatype of a cell) and within a one step linear parse of the matrix, assigns a connected component
to each cell. 
%a well-know algorithm in computer vision. 
%to perform connected component detection (i.e. partition cells per regions /tables).

%\begin{scriptsize}
\begin{tiny}
\begin{algorithm}[h]
%\dontprintsemicolon
\SetKwInOut{Input}{input}
\SetKwInOut{Output}{output}
\Input{type(i,j): a function returning the datatype of each cell }
\Output{cc(i,j) : a function returning the connected component of each cell }
%\lnl{ln:init}
%$\F \leftarrow \K$
%\lnl{ln:mainLoop}
\ForEach{$0<i<n$} {
  \ForEach{$0<j<m$} {
    \lIf{$cc(i-1,j) \neq null$} {$cc(i,j) \leftarrow cc(i-1,j)$}
    
    \lElse{$cc(i-1,j-1) \neq null$} {$cc(i,j) \leftarrow cc(i-1,j-1)$}
    
    \lElseIf{$cc(i,j-1) \neq null$} {$cc(i,j) \leftarrow cc(i,j-1)$}
    
    \lElseIf{$cc(i-1,j+1) \neq null$} {$cc(i,j) \leftarrow cc(i-1,j+1)$}
    
    \ElseIf{$type(i,j) \neq void$} {$cc(i,j) \leftarrow new~ConnetedComponent()$}
   }
 }
 \caption{\textsc{Table Detection with Connected Component}}
\label{algo:cc}
\end{algorithm}
\end{tiny}

Algorithm \ref{algo:cc}  allows us to partition the spreadsheet into regions. 
We then use more sophisticated  computer vision approaches, such as  morphologic
transformation \cite{Haralick1992} and erode / dilate functions \cite{Kong1996} to refine
the result of the connected component detection: remove too small
connected components,
%PV: je ne comprend pas "remove to small"
merge connected components that have been splitted due to a single void line, etc...

\begin{figure}[h]
     \includegraphics[width=8.7cm]{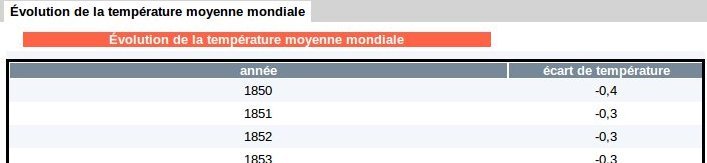}
     \caption{Table detection in an Excel file}
     \label{fig:table}
\end{figure}

In the graphical interface (see Figure \ref{fig:table}), the detected tables are drawn within a frame.

To decide whether data are line- or column-oriented, we exploit the following idea: if data are presented in
lines, the datatypes of cells for each line are homogeneous, but the datatypes of cells for each column may 
be heterogeneous. 
We then compute the discrepancy in terms of cell datatypes for each line (\ref{discr_lines}) and for 
each column (\ref{discr_columns}). If (\ref{discr_lines}) > (\ref{discr_columns}), 
then the metadata are probably on the first lines, or on the first columns otherwise.

%$$\sum_{0<i<n} \max_{t\in \{string, int, \ldots\}} line_i^{type}$$
%$$\sum_{0<i<n} (\max_{\forall type} (line_i^{type}))$$
\begin{small}
\begin{equation}\label{discr_lines}
%discrepancy(lines) \leftarrow 
\sum_{0<i<n} (\max_{t\in \{string, int, \ldots\}} (\sum_{0<j<m} (type_{[i,j]}= t )))
\end{equation}

\begin{equation}\label{discr_columns}
 %discrepancy(columns) \leftarrow 
 \sum_{0<j<m} (\max_{t\in \{string, int, \ldots\}} (\sum_{0<i<n} (type_{[i,j]}= t )))
\end{equation}
\end{small}

The end of the process relies on machine learning \cite{Mitchell1997}. 
%Based on these discrepancy measures and additional criterions (such as datatype of a cell, 
 Using past experience and based on several criterions:  the discrepancy measures, the datatype of a cell,
 the data type of the neighborhood of a cell, 
WebSmatch detects each important component in the spreadsheet file such as: titles, comments, table data, table header (see Figure \ref{fig:table}).
Machine learning is able to capture several spreadsheet users habits,
such as: ``cells on the very first line of a connected component,
having the string datatype 
and bellow cells having a numeric datatype are often metadata'' or
``cells having the string datatype and void neighborhood  and behind a
table often are a title''.
The important feature is that such rules have not been designed by the user, but observed on several documents. They can be updated when 
new spreadsheets are performed by the user.

\subsection*{Matching}
WebSmatch relies on YAM++ \cite{YAM++} to perform the matching task.
YAM++ combines 14 different matching techniques,  divided in 3 main groups: string matchers, dictionary and thesaurus matchers based on Wordnet\footnote{\url{http://wordnet.princeton.edu/}} and instance-based matchers.
Instance-based matcher is the generic name for matchers, which deals both with metadata and data. 
Such matchers are very useful when the column names are not informational enough, which is often
the case in public data. 
The instance-based matcher implemented in YAM++ is very powerful and one of the main
reasons for YAM++ excellent results at the 2011 competition of the Ontology
Alignment Evaluation Initiative (\url{http://oaei.ontologymatching.org}:
first position at the Conference track and second position at the
Benchmark track \cite{YAM-OAEI}.

\begin{figure}[h]
     \includegraphics[width=8cm]{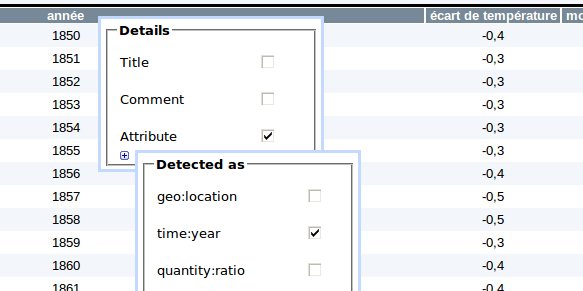}
     \caption{Matching sources with  DSPL concepts}
     \label{fig:TimeYear}
\end{figure}

\begin{figure}[h]
     \includegraphics[width=8cm]{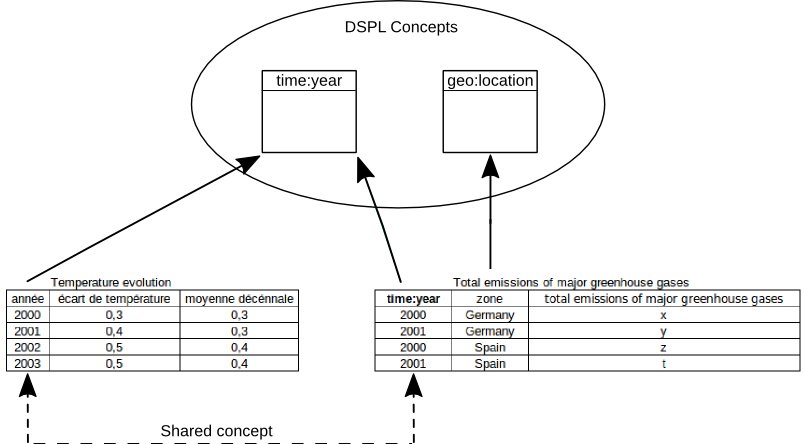}
     \caption{Result of integration}
     \label{fig:integration}
\end{figure}

Figure \ref{fig:TimeYear} is a zoom of Figure \ref{fig:table} on  the
cell ``année'' (i.e. year in french),  which has been previously
detected as metadata. This cell is detected as "time:year" concept
by applying the instance-based matcher on its data collection $\{1990,1991,\ldots\}$.
Figure \ref{fig:integration} depicts all the discovered matches over
the two files of the scenario and the DSPL concepts we previously imported into the tool. 

Notice that the line of the second spreadsheet (Figure  \ref{fig:xls}) contains a line within a collection of years but with a void cell as first column. Despite it is void, this cell is detected by WebSmatch to be a metadata. Indeed,
it is at the first line and first column of the detected table and our
machine learning algorithm detects the metadata to be 
placed in the first column. By applying the instance-based matcher, WebSmatch suggests this cell to be labelled with the "time:year" concept.

\subsection*{Clustering}
%\textbf{Attention ce paragraphe provient d'une précédente version du papier, les captures d'écrans
%ne correspondent pas à l'exemple de ce papier ! }
Based on the semantic links discovered by the matchers between documents, WebSmatch
automatically clusters the set of documents. It first computes a distance between
each pair of documents. More formally, we build a bipartite graph, where nodes are
attributes from the documents 
%from left are the attribute (meta-data) of document 1, nodes from
%right are the attributes of document 2, 
and edges are the matches discovered by the matching services, the
weights over edges are labelled
%PV: manque un mot entre "are" et "by"
by the confidence value of the discovered matches.
From this weighted  bipartite graph, we compute the maximum matching and normalize 
it by dividing it by the minimum numbers of attributes between the two documents.% 1 and 2.

\begin{figure}[h]
\hspace{-0.7cm}\includegraphics[width=9cm]{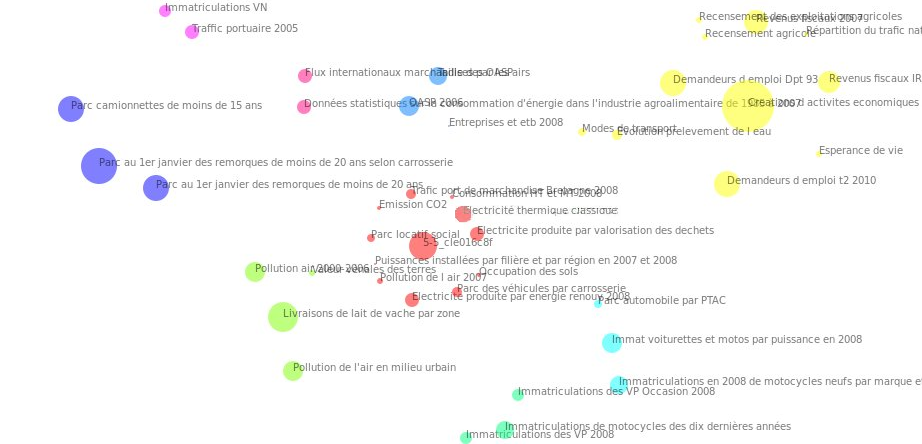}
\caption{The cluster}
\label{fig:cluster}
\end{figure}

From these distances between documents, we propose a minimum energy graph model (adapted from \cite{linlog}), 
which represents the cluster structure based  on repulsion between documents. 
Figure \ref{fig:cluster} illustrates the result of the clustering service after adding a given number of documents:
each cluster is drawn in a different color, documents are in the same cluster if and only if they share some
semantics links. Documents have different diameters: the larger is the diameter, the more representative of the cluster is the document.

The clustering service provides an automatic way to classify documents
in several categories. This is a very interesting feature in the Data Publica application, where the number of sources is huge ($>12.000$).
Finally, it is able to perform some recommendation, by suggesting to the user documents
related to those she is currently interested in.
%For instance, if interested in the "Puissances installees, productions d'electricite et de chaleur, combustibles ...." document, WebSmatch will suggest the user to look at the document labelled "Electricitite thermique classique", another important document (since with a large radius) of the same cluster, and suggests to the user to switch to the matching service between these two documents .

\subsection*{Visualization} 

By detecting the blank cell, we are able to convert the bi-dimensionnal table from the 
initial spreadsheet (Figure  \ref{fig:xls}) into a  classical (SQL-like) flat table (Figure \ref{fig:integration}).
Thanks to the matching process, we are also able to identify concepts (from DSPL) over the data sources
and to detect common attributes in order to produce integrated data.

At this step, we have distinguished data and metadata from the initial
Excel files, and flatted bi-dimensionnal tables. 
%Remi: je ne comprend pas la fin de cette phrase; tu veux dire "a
%cleaned table?" on ne parle jamais de relation. que veut dire
%"dispose" ici?
%dispose from cleaned  relation.
We can easily generate an XML file describing the metadata (title, header, concepts) and the 
\url{.csv} files containing the data to fit the strict DSPL input format. 
As a result, we can take advantage of the powerful capabilities of Google Data Explorer in terms of visualization
%Modif 27 feb
or load the structured data into Data Publica's database as shown in Section \ref{sec:scenario}.

\section{Related Work}
\label{sec:related}

%\textbf{Les papiers que le reviewer de EDBT nous a demandé d'ajouter}
In terms of metadata extraction, 
the problem of identifying charts in documents using machine learning techniques has been widely studied over the last decade. 
In \cite{Huang2007}, the authors propose a method to automatically detect bar-charts and pie-charts, using computer vision
techniques and instance-based learning. 
The approach developed in \cite{Classifying2007}  relies on a
multiclass Support Vector Machine, as machine learning classifier. It
is able to identify more kinds of  charts, namely  bar-charts,
curve-plots, pie-charts, scatter-plots and surface-plots.
%
%in a document, using image segmentation based on colors, and a
%multi-class Support Vector Machine, as Machine Learning classifiers. In \cite{Huang2007}, a 
More generally,  \cite{Blostein2000} presents a survey of extraction techniques of diagrams in complex documents,
such as scanned documents.  

All these techniques allow recognition of charts, thus much complex shapes than tables. But, in our case our problem
is not only to decide whether a table is present or not in the document, but to provide precise coordinates of all tables in
the document. \\

%\textbf{Google tools}

%\begin{itemize}
%\item Situer WebSmatch vis à vis de Google Refine.
%\item Situer aussi l'API de DataPublica vis à vis de Google Fusion Tables  (API)  / Google Public Data Explorer et 
%Google Spreadsheets (visualisation) ? -> Si, oui demander à Christian. 
%\end{itemize}
%Google Spreadsheets \url{http://www.google.com/google-d-s/spreadsheets/}
%Google Fusion Tables \url{http://www.google.com/fusiontables/}: Upload data tables from spreadsheets or CSV files, even KML. Developers can use the Fusion Tables API to insert, update, delete and query data programmatically. You can export your data as CSV or KML too. 

Google Refine (\url{code.google.com/p/google-refine/}) is a powerful tool to perform data cleaning. 
It helps the user to deal with messy data, by discovering
inconsistencies. For instance,  
it allows string transformation to avoid the same entity, spelled in two different ways to be considered as two
different entities. 
Google Refine also allows data augmentation using external web services
or named-entity recognition based on the FreeBase social database (\url{http://www.freebase.com}).
Using the ``Add column based on a URL fetched on column'', the user can add extra columns to her document. 
Nevertheless, she needs to know precisely which service to call and its complete syntax. 

The major drawback of Google Refine when dealing with Excel files is the strict assumptions made over the input spreadsheet.
Excel files need to have exactly one table per sheet and all attributes have to be in column and at the first line of the sheet (or 
the number of header lines have to be explicitly mentioned). 
WebSmatch's metadata extraction service is thus a mandatory step to
use Google Refine on documents such as those 
published by french administrations and crawled  by DataPublica. 

Another cooperation between WebSmatch and Google Refine deals with data augmentation. 
Thanks to its matching capabilities, WebSmatch is able to tag the
first column of a document (Figure \ref{fig:xls})
with DSPL concepts (namely geo:location). Geo-encoding such column may
then be done automatically, without any involvement of the user. 

%WebSmatch is not a concurrent of Google Refine 

\section{Conclusion}
\label{sec:conclu}

In this paper, we described WebSmatch, a flexible environment for Web
data integration, based
on a real data integration scenario over public data
from Data Publica. We chose a typical scenario 
that involves problems not solved by currents tools: poorly
structured input data sources (XLS files) and rich visualization of
integrated data. WebSmatch supports the full process of importing,
refining and integrating data sources and uses third party tools
%(from Google Data Explorer) 
for high quality visualization and data delivery. 
%The demonstration will provide additional scenarios, involving  more
%data sources and including data source recommendation using the clustering service.
A video playing the whole motivation example is available at \url{http://websmatch.gforge.inria.fr}.
Furthermore, it can be played with a test account at the same url. 
%PV: manque un lien
.

\bibliographystyle{abbrv}
\bibliography{OpenData}

\end{document}